\font\caps=cmcsc10 at 12pt
\font\eightrm=cmr8 at 8pt
\documentclass{oxarticle}
\usepackage{amsmath, amssymb}
\newcommand{\tr}{{transformation}}
\newcommand{\DI}{{ Dirac Irreducible Superspin $\fr{1}{2}$   Massive  Multiplet}}

\newcommand{\cA}{{\cal A}}

\newcommand{\CC}{Complex Conjugate}

\newcommand{\PB}{BRST Poisson Bracket}

\newcommand{\LT}{\LaTeX}

\newcommand{\bt}{\begin{tabular}{c}}
\newcommand{\et}{\end{tabular}}

\newcommand{\eb}{\ee\be } 
\newcommand{\ebp}{\rt.\ee\be\lt.} 
\newcommand{\bmat}{\lt ( \begin{array} }
\newcommand{\emat}{  \end{array} \rt )}

\newcommand{\ovW}{{\ov W}}

\newcommand{\cP}{{\cal P}}

\newcommand{\oB}{{\ov B}}

\newcommand{\ovV}{{\ov V}}
\newcommand{\ovD}{{\ov D}}

\newcommand{\oJ}{{\ov J}}

\newcommand{\oE}{{\ov E}}

\newcommand{\oU}{{\ov U}}

\newcommand{\ovG}{{\ov G}}
\newcommand{\oK}{{\ov K}}

\newcommand{\ED}{\end{document}}

\newcommand{\og}{{\ov g}}

\newcommand{\oC}{{\ov C}}

\renewcommand{\a}{\alpha}	
\renewcommand{\b}{\beta}
\newcommand{\g}{\gamma}
\renewcommand{\d}{\delta}

\newcommand{\h}{\eta}

\newcommand{\x}{\xi}

\newcommand{\f}{\phi}
\renewcommand{\c}{\chi}

\newcommand{\w}{\omega}

\newcommand{\D}{\Delta}

\newcommand{\X}{\Xi}
	
\renewcommand{\S}{\Sigma}
\newcommand{\U}{\Upsilon}

\newcommand{\W}{\Omega}
\newcommand{\la}{\label}
\newcommand{\ci}{\cite}

\newcommand{\ds}{\documentstyle}	
\newcommand{\fr}{\frac}

\newcommand{\pa}{\partial}
\newcommand{\ov}{\overline}
\newcommand{\be}{\begin{equation}}
\newcommand{\ee}{\end{equation}}
\newcommand{\ba}{\begin{array}} 
\newcommand{\ea}{\end{array}}
\newcommand{\bea}{\begin{eqnarray}}
\newcommand{\eea}{\end{eqnarray}}
\newcommand{\ra}{\rightarrow}

\newcommand{\lt}{\left}
\newcommand{\rt}{\right}

\newcommand{\ben}{\begin{enumerate}}
\newcommand{\een}{\end{enumerate}}
\newcommand{\bitem}{\begin{itemize}}
\newcommand{\eitem}{\end{itemize}}

\newcounter{orange} 
\addtocounter{orange}{1}

\newcommand{\articlenumber}{\LT 4350.FirstDISHpapersubmission}

\proofmodefalse
\usepackage{color} 

\begin{document}

\begin{center}

%\vspace*{1in}
{  \huge {  \huge
An Irreducible Massive Superspin One Half Action Built From the Chiral Dotted Spinor Superfield
\\[1cm] }}
%\articlenumber\\[1cm]\
%

%\vspace*{.1in}
%{    J. A. Dixon}
%

%\today

\renewcommand{\thefootnote}{\fnsymbol{footnote}}
%\footnotetext[1]{~here we have a footnote.}\renewcommand{\thefootnote}\arabicfootnote}}

%sign
{\caps John A. Dixon\footnote{dixon@maths.ox.ac.uk; \;jadixg@gmail.com}\\
Mathematical Institute,\\ Oxford University \\ Oxford, England}\\[1cm] 
%\vfill

%\today

{\bf Abstract}
\end{center}
\normalsize
%\LARGE
%\large
%\normalsize
Although the  chiral dotted spinor superfield  should describe a Massive Superspin One Half multiplet, it has not been obvious how to derive this from an action. In this paper this is done by including  a  chiral undotted spinor superfield, finding the BRST transformations that govern both of these, and then finding the action as an invariant of the transformations. 
It turns out that both kinds of spinor superfields are needed.  Moreover, the BRST transformations for the two kinds of chiral spinor superfields are generated from each other by a special involution that exchanges Grassmann odd (even) sources with Grassmann even (odd) fields.

\large

%\begin{tabular}{c}\\ \hline \hspace{6in}\\
%\end{tabular}

%\cite{Gatesfundremains}
%\cite{Calkinsthinkdiff}
%\cite{Buchbindersupone}
%\cite{Buchbindersupthree}
%\cite{KBbook}
%\cite{GatesKoutmassless}
%\cite{Gateslinmassdyn}
%\twocolumn
\large

{\bf \theorange}.\;\addtocounter{orange}{1}
 In some suitable limit for the superstring \ci{newwest}, the massive modes    might be described by massive supersymmetric actions, coupled in some 
way to each other. As noted by \ci{Zinoviev:2007js} and as emphasized rather recently by Gates and Koutrolikos 
\cite{{Gateslinmassdyn}}, massive supersymmetric theories possess a rich off shell structure and there is still much to learn about them.  
In  \cite{Gatesfundremains} the authors posed the `Off Shell Susy Problem' in a simple and general way, and pointed out that the answer is not likely to be simple, but that it is probably important. This has generated the adinkra approach which is making progress on this complicated problem \cite{Calkinsthinkdiff}.

Assembling interacting actions is a significant problem when one has only  `on-shell closure' 
since this necessarily implies some particular action of course, and that makes it tricky to generalize the action to include other couplings.
It is usually taken for granted that the best way to approach the problem of generating actions and couplings is to look for auxiliary fields, and actions expressed in terms of superfields, so that the SUSY algebra closes on shell and the SUSY transformations are then obvious from superspace theory. 
Progress using these ideas has been reported in \ci{Buchbindersupone,Buchbindersupthree} 
and \ci{GatesKoutmassless}.
Of course, the general problem, as noted in 
\cite{Gatesfundremains},  is complicated by the existence of other symmetries, such as those that must exist in the Standard Model of particle theory.

%\cite{Gatesfundremains}
%\cite{Calkinsthinkdiff}
%\cite{Buchbindersupone}
%\cite{Buchbindersupthree}
%\cite{KBbook}
%\cite{GatesKoutmassless}
%\cite{Gateslinmassdyn}

{\bf \theorange}.\;\addtocounter{orange}{1}
The intent in this paper, and its sequel, is to show that the BRST approach, using cohomology, offers a different approach to some of these problems. 
The present paper will illustrate some of the issues here by constructing a massive superspin $\fr{1}{2}$ action out of superfields with   spin $\fr{1}{2}$.  
We will not use superfields here, except at the start, but the algebra is closed in the sense that the BRST operator is nilpotent.  The nilpotence of the BRST operator 
arises as though the auxiliaries have been integrated out. This can happen even when no auxiliaries exist\footnote{This is  probably the case for the present action, and also for  10-D Super Yang Mills theory 
\cite{Calkinsthinkdiff,dix10D}, for example.  
This feature is related to the Batalin Vilkovisky method, see for example \ci{Weinberg} for a simple exposition of the latter.}.
If there is a nilpotent BRST operator,  one can use the spectral sequence to discover the cohomology  \cite{Dixonspecseq}.  The cohomology then points out where there are new invariants.

This BRST approach singles out the physical fields, and any remaining auxiliary fields get eliminated from consideration early on in the analysis \cite{Dixonspecseq}.  So the BRST approach generates a different set of insights and problems, and it is not simple to sort out the relationship between the superfield  approach and the BRST cohomology approach.
They are complementary.

{\bf \theorange}.\;\addtocounter{orange}{1}
For a number of reasons to do with BRST cohomology\footnote{It has been evident for a long time
 \cite{Dixonholes,Dixonchiralcohom,DixonRahm,Dixonjumps}
 that the BRST cohomology of the   chiral scalar superfield ${\widehat A}$ couples naturally to a chiral dotted spinor superfield ${\widehat\f}^{\dot \a} $.  The simplest example is $\int d^6 z  {\widehat\f}^{\dot \a} {\widehat A} \oC_{\dot \a}$. Here $ \oC_{\dot \a}$ is a spacetime constant supersymmetry ghost.  However the superfield ${\widehat\f}^{\dot \a} $ here needs further constraints, which is the progress reported in this paper.}, it is of some interest to construct a well-behaved  action starting with a chiral dotted spinor superfield $ {\widehat  \f}_{ \dot \a}
$.  
 Chirality means that $ \ovD_{\dot \b}{\widehat  \f}_{ \dot \a}=0
$.  It is well known \ci{KBbook} that this superfield can be subjected to a `reality constraint' in addition to chirality, and that on-shell it should represent superspin $\fr{1}{2}$.  But the problem is that no action has been found, until now, that is consistent with both the chirality and the reality constraints and which  then yields superspin 
$\fr{1}{2}$.

{\bf \theorange}.\;\addtocounter{orange}{1}
Trying to write down an action for 
a chiral dotted spinor superfield $ {\widehat  \f}_{ \dot \a}$ 
 meets a problem right at the start, because the most obvious action is

\be
\cA= 
\int d^8 z 
{\widehat \f}_{ \dot \a} \pa^{\a \dot \a}
{\widehat {\ov\f}}_{  \a}
+m^2 \int d^6 z
 {\widehat \f}_{}^{\dot \a}
 {\widehat \f}_{\dot \a}
+m^2 \int d^6 \ov z
 {\widehat {\ov\f}}_{}^{  \a}
 {\widehat {\ov \f}}_{ \a}
\ee
but this immediately leads to higher derivative equations of motion, and there are tachyons in the spectrum too:
\be
\lt ( \Box^2 - m^4 \rt ) 
{\widehat \f}_{ \dot \a}  =
\lt ( \Box - m^2 \rt ) 
\lt ( \Box + m^2 \rt ) 
{\widehat \f}_{ \dot \a}  =0
\ee
This is certainly not a promising start for a model that is supposed to be phenomenologically viable.

{\bf \theorange}.\;\addtocounter{orange}{1}
The 
chiral undotted  spinor superfield ${\widehat {\ov\c}}_{  \a}$ should also yield massive superspin $\fr{1}{2}$ on shell  \ci{KBbook} and this is also puzzling. Here  chirality means that $ D_{ \b}
{\widehat  \c}_{ \dot \a}=0
$. ${\widehat {\ov\c}}_{  \a}$ does appear in gauge theory\ci{vecsupref}, but as a massive matter representation it poses a difficulty, because the only known way to give it mass, until the present action, involves the spontaneous breaking of gauge symmetry together with the introduction of Higgs scalars.  Such a method to construct a massive representation of superspin $\fr{1}{2}$ is clearly not irreducible, because it mixes ${\widehat \c}_{\dot \a}$ with the components of chiral scalar Higgs multiplets.

{\bf \theorange}.\;\addtocounter{orange}{1}
Here we also want to add another feature, which is phase invariance, corresponding to some   conserved quantity like Lepton or Baryon number.  So we add a chirality index $L,R$ to the superfields.  Then 
 the chirality constraints have the form:
\be
\ovD_{\dot \b}
{\widehat \f}_{L\dot \a}
=
\ovD_{\dot \b}
{\widehat \f}_{R\dot \a}
=
D_{ \b}
{\widehat \c}_{L\dot \a}
=
D_{ \b}
{\widehat \c}_{R\dot \a}
=
0
\la{ccts}
\ee
So the \CC s satisfy:
\be
D_{\a}
{\widehat {\ov\f}}_{L\b}
=
D_{\a}
{\widehat {\ov\f}}_{R\b}
=
\ovD_{\dot \a}
{\widehat {\ov\c}}_{L\b}
=
\ovD_{\dot \a}
{\widehat {\ov\c}}_{R\b}
=
0
\la{cctscc}
\ee
Next, in order to get an irreducible representation of supersymmetry, along with a phase invariance, along the lines of the textbook 
\ci{KBbook}, we want to also impose the additional `reality constraints':
\be
D_{\a} 
{\widehat \f}_{L\dot \a}
=
 \ovD_{\dot\a} 
{\widehat {\ov\f}}_{R\a}
;\ovD^{\dot \a}  
{\widehat \c}_{R\dot \a}
=
 D^{ \a}  
{\widehat {\ov\c}}_{L \a}
\la{reccts}
\ee

These are designed so that there is a global U(1) phase invariance that is conserved by the action.

{\bf \theorange}.\;\addtocounter{orange}{1}
\la{pbsection}
In the context of BRST \ci{Zinnarticle, Weinberg}, a theory is defined by its BRST transformations, which can be derived from its \PB.   Here is the \PB\ of the present theory:

\be
\cP_{\rm Total}[\cA]=
\cP_{\rm \c}[\cA]+
\cP_{\rm \f}[\cA]+
\cP_{\rm SUSY}[\cA]
\ee

\be
\cP_{\rm \c}[\cA]=
\int d^4 x \lt \{
\fr{\d \cA}{ \d U_{R\dot \a}}
\fr{\d \cA}{ \d \c_L^{\dot \a}}
+
\fr{\d \cA}{ \d U_{L\dot \a}}
\fr{\d \cA}{ \d \c_R^{\dot \a}}
\ebp
+
\fr{\d \cA}{ \d \ov\W_{\a\dot \a}}
\fr{\d \cA}{ \d V^{\a\dot \a}}
+
\fr{\d \cA}{ \d \ov \X}
\fr{\d \cA}{ \d B }
\ebp+\fr{\d \cA}{ \d \oK}
\fr{\d \cA}{ \d  \w}
+\fr{\d \cA}{ \d \oJ }
\fr{\d \cA}{ \d \h}
+\fr{\d \cA}{ \d \ov\D }
\fr{\d \cA}{ \d L}
+*
\rt \}
\ee

\be
\cP_{\rm \f}[\cA]=
\int d^4 x \lt \{
\fr{\d \cA}{ \d Z_L^{\dot \a}}
\fr{\d \cA}{ \d \f_{R\dot \a}}
+
\fr{\d \cA}{ \d Z_R^{\dot \a}}
\fr{\d \cA}{ \d \f_{L\dot \a}}
\ebp
+
\fr{\d \cA}{ \d \ov\S^{\a\dot \a}}
\fr{\d \cA}{ \d W_{\a\dot \a}}
+
\fr{\d \cA}{ \d \ov \U}
\fr{\d \cA}{ \d E }
\ebp
+\fr{\d \cA}{ \d \ov J'}
\fr{\d \cA}{ \d \eta' }
+\fr{\d \cA}{ \d \ov K' }
\fr{\d \cA}{ \d \w'}
+\fr{\d \cA}{ \d \ov \D' }
\fr{\d \cA}{ \d L'}+*
\rt \}
\la{phipb}
\ee

\be
\cP_{\rm SUSY}[\cA]
=
\fr{\pa \cA}{ \pa h_{\a\dot \a} }
\fr{\pa \cA}{ \pa \x^{\a\dot \a} }
\la{susypb}
\ee

As emphasized above, in  this paper we do not try to keep manifest supersymmetry.  We decompose the superfields into components and look for nilpotent BRST transformations, which then generate the action. All the above fields and sources are components, not superfields.

We note that each term in the above, such as the first one $\fr{\d \cA}{ \d U_{R\dot \a}}
\fr{\d \cA}{ \d 
\c_L^{\dot \a}}
$, contains one Zinn source derivative (here it is $U_{R\dot \a}$)
 and one Field 
derivative (here it is $\c_L^{\dot \a}$), and one of them is Grassmann even ($U_{R\dot \a}$ here), and the other odd ($\c_L^{\dot \a}$ here).
We will discuss the meaning of the symbols more fully below after equation 
(\ref{kinchi}),
where we write down the action.

The Action $\cA$ of the theory contains two parts:
\be \cA = 
\cA_{\rm Zinn}+ 
\cA_{\rm Fields} 
\ee

To start with, one must find an action 
$\cA_{\rm Zinn}$ such that the related \PB\ vanishes identically.  This action generates the transformations.
  We can define a sort of square root of the \PB, by
\be
 \d_{\rm First}= 
\d_{\rm Fields}
+
\d_{\rm Zinns}
\ee
where 
\be
\d_{\rm Fields}
=
\sum_i
\int d^4 x  
\fr{\d \cA_{\rm Zinn}}{ \d 
{\rm Zinn}_i
}
\fr{\d }{ \d 
{\rm Field}_i}
\eb = 
\int d^4 x \lt \{
\fr{\d \cA}{ \d U_{R\dot \a}}
\fr{\d }{ \d \c_L^{\dot \a}}
+
\fr{\d \cA}{ \d U_{L\dot \a}}
\fr{\d }{ \d \c_R^{\dot \a}}
+\cdots
\rt )
+*\ee
and
where 
\be
\d_{\rm Zinns}
=
\sum_i
\int d^4 x  
\fr{\d  \cA_{\rm Zinn}}{ \d {\rm Field}_i}
\fr{\d }{ \d 
{\rm Zinn}_i}
\eb = 
\int d^4 x \lt \{
\fr{\d  \cA_{\rm Zinn}}{ \d \c_L^{\dot \a}}
\fr{\d }{ \d U_{R\dot \a}}
+
\fr{\d  \cA_{\rm Zinn}}{ \d \c_R^{\dot \a}}
\fr{\d }{ \d U_{L\dot \a}}
+\cdots
\rt )+*
\ee
If it is true that
\be
\cP[ \cA_{\rm Zinn}]=0
\ee
then\footnote{In the present case this further reduces to suboperators:
$
\d_{\rm Fields}=
\d_{\rm \c\;Fields}+
\d_{\rm \f\; Fields}
$ and $\d_{\rm Zinns}=
\d_{\rm \c\; Zinns}+
\d_{\rm \f\; Zinns}
$
and all these suboperators are nilpotent or they anticommute:
$\d_{\rm Fields}^2
=\d_{\rm Zinns}^2
=
\lt \{
\d_{\rm Fields}
,\d_{\rm Zinns}
\rt\}
=\d_{\rm \c\;Fields}^2=
\d_{\rm \f\; Fields}^2=
\d_{\rm \c\; Zinns}^2
=
\d_{\rm \f\; Zinns}^2
=0$
etc.
} it follows that
\be
 \d_{\rm First}^2
=0.
\ee

Next one looks for an action that satisfies the invariance identity
\be
\d_{\rm Fields}
\cA_{\rm Fields} =0
\ee
where the expression 
$ \cA_{\rm Fields}$ is in the cohomology space of $\d_{\rm Fields}$ and depends only on Fields and not on Zinns.
Since $\cA_{\rm Fields}$ depends only on Fields, it follows that 
\be \cP[\cA] = 
 \cP[\cA_{\rm Fields}] = 
0
\ee
The expression $\cP[\cA_{\rm Fields}] $ is trivially zero, because it contains no Zinns, and each term of the \PB\ contains one Zinn.

{\bf \theorange}.
\la{kineticaction}
\;\addtocounter{orange}{1}
As we pointed out above, a theory can be constructed from 
\ben
\item
A Form of Poisson Bracket, which describes the Fields, Zinns and their pairing;
\item
A Zinn Action which describes the transformations of the Fields, and, through the \PB, also the transformations of the Zinns;
\item
A field action which is invariant under the field Transformations. 
\een
We will start with the field action, since it is shorter. 

{\bf \theorange}.\;\addtocounter{orange}{1}
 Here is the action \ci{vecsupref} that arises from the chiral undotted   $\c$-type   superfields referred to above:
\[
\cA_{\rm Kinetic \;\c}
=
 \int d^4 x \; \lt \{
\c_{L}^{\dot \a}
\pa_{\a\dot\a}
\ov\c_{L}^{ \a}
+
\c_{R}^{\dot \a}
\pa_{\a\dot\a}
\ov\c_{R}^{ \a}
\rt.
\]
\be
\lt.
+
G_{\dot \a\dot \b}
\ovG^{\dot \a\dot \b}
-2
B \oB
\rt \}
\la{kinchi}
\ee

where
\[
G^{\dot \a \dot \b} 
=
G^{(\dot \a \dot \b)} 
=
\fr{1}{2}
\lt (
\pa_{\g}^{\dot \a} 
V^{\g \dot \b} 
+
\pa_{\g}^{\dot \b} V^{\g \dot \a} 
\rt )
\]
\be 
\ovG^{\a \b} 
=
\ovG^{(\a \b)} 
=
\fr{1}{2}
\lt (
\pa_{\dot\g}^{ \a} \ovV^{\dot\g \b} 
+
\pa_{\dot\g}^{ \b} \ovV^{\dot\g \a} 
\rt )
\la{defofG}
\ee
and  also 
\[
\ovG^{(\dot \a \dot \b)} 
=
\fr{1}{2}
\lt (
\pa_{\g}^{\dot \a} \ovV^{\g \dot \b} 
+
\pa_{\g}^{\dot \b} \ovV^{\g \dot \a} 
\rt )
\]
\be
G^{(\a \b)} 
=
\fr{1}{2}
\lt (
\pa_{\dot\g}^{ \a} V^{\dot\g \b} 
+
\pa_{\dot\g}^{ \b} V^{\dot\g \a} 
\rt )
\la{defofGbar}
\ee

In the above, $\c_{L}^{\dot \a}$ and $\c_{R}^{\dot \a}$  are two-component Weyl spinors, $B$ is a complex scalar (it turns out to be an auxiliary field), and $G$ is a complex field strength made from a complex vector field 
$V^{\g \dot \b} 
$.
More explicitly we have
\be
 \int d^4 x \; \lt \{
G_{\dot \a\dot \b}
\ov G_{}^{\dot \a\dot \b}
\rt \}
\eb
=
\int d^4 x 
 V^{\a \dot \a}
\lt (
\Box
\ovV_{\a \dot \a} 
+
\fr{1}{2}
\pa_{\a \dot \a} \pa_{\g \dot \d} \ovV^{\g  \dot \d} 
\rt )
\ee

{\bf \theorange}.\;\addtocounter{orange}{1}
Here is the action that arises from the chiral dotted   $\f$-type   superfields referred to above:
\[
\cA_{\rm Kinetic \;\f}
= \int d^4 x \; \lt \{
\f_{L}^{\dot \a}
\pa_{\a\dot\a}
\ov\f_{L}^{ \a}
+
\f_{R}^{\dot \a}
\pa_{\a\dot\a}
\ov\f_{R}^{ \a}
\rt.
\]
\[
+
 W_{\a\dot \a}
\ovW^{\a\dot \a}
-\fr{1}{2}
E \Box \oE 
+
 \fr{1}{2}  
\ov \eta'  \lt (
 \f_L^{\dot \d}
\oC_{\dot \d}
+
\ov\f_R^{\d}
C_{ \d}
\rt )
\]
\be
\lt.
+\fr{1}{2}
 \eta'  \lt (
 \ov\f_L^{  \d}
C_{\d}
+
 \f_R^{\dot\d}
\oC_{\dot \d}
\rt )
\rt \}
\la{actphi}\ee

In the above, $\f_{L}^{\dot \a}$ and $\f_{R}^{\dot \a}$  are two-component Weyl spinors, $E$ is a complex scalar 
field, $W_{\a \dot \a}$ is a complex vector field (it turns out to be an auxiliary field);
$\h'$ is a complex ghost antifield 
and $C_{\a}$ is a constant Weyl spinor ghost corresponding to the rigid SUSY transformations.

The above kinetic actions are determined as invariants of the field transformations.
Actually here these split into two:
\be
\d_{\rm \c\;Fields}
\cA_{\rm Kinetic \;\c}
=0
\ee
\be
\d_{\rm \f\;Fields}
\cA_{\rm Kinetic \;\f}
=0
\ee

As mentioned above, these transformations $\d_{\rm Fields}
$ follow from the Zinn actions.
Now we present these Zinn actions and explain where they come from.

{\bf \theorange}.\;\addtocounter{orange}{1}
\la{zinnforchi}
 First we have the Zinn Action  for the $\c$ sector, which follows from the textbook treatment of gauged supersymmetry
 \ci{vecsupref}.  We can write it so that the derivatives with respect to Zinns are easy to see, as follows:
\be
 \cA_{\rm Zinn\;\c}
\la{chiwithzinnsout}
\eb
=\int d^4 x
\lt\{ U_{R  \dot \a}
 \lt (
B \oC^{\dot \a}
+
G^{(\dot \a \dot \b)} 
 \oC_{\dot \b}
+ \x \cdot \pa\;
 \c_L^{\dot  \a}
\rt)
\rt.
\eb
+  U_{L  \dot \a}
 \lt (
- \oB \oC^{\dot \a}
+
\ov G^{(\dot \a \dot \b)} 
 \oC_{\dot \b}
+ \x \cdot \pa\;
 \c_R^{\dot  \a}
\rt)
\ee
\be
+
  \Xi
\lt (
\fr{1}{2}
 \pa^{\a  \dot \a} 
\ov\c_{L \a} \oC_{\dot\a}
-\fr{1}{2}
 \pa^{\a  \dot \a} 
 \c_{R \dot\a}  C_{\a}
+ \x \cdot \pa\;
\oB
\rt )
\ee

\be
+
\W_{ \a\dot \a}
\lt (
\pa^{ \a\dot \a}
\ov \w +  \c_R^{\dot \a}
C^{\a}
+
 \ov\c_L^{\a}\oC^{\dot \a} 
+ \x \cdot \pa\;
\ovV ^{\a\dot \a}
\rt )
\la{gaugetrans}
\ee
 
\be
+K
\lt (
 \ovV^{\b\dot \b}
C_{\b} \oC_{\dot \b}
+ \x \cdot \pa\;
\ov \w
\rt )
\eb+  
\oJ  
\lt (
L + \x \cdot \pa\;
\h
\rt )
\lt.
+ 
\ov \D 
\lt (
C_{\b} \oC_{\dot \b}
\pa^{\b \dot \b}
\h
+ \x \cdot \pa\;
L
\rt )
\rt \}
\ee

In the above we have some more notation in addition to that noted after equation (\ref{defofGbar}).
In the above, $\w$ is a complex anticommuting Faddeev Popov Ghost,
$\h$   is  the corresponding complex Faddeev Popov
 antighost, and $L$ is a commuting scalar field which is auxiliary and useful for dealing with the gauge fixing term, as shown below.
The Zinn sources $
U_{R\dot \a}$ etc. are conjugate under the \PB\ to the fields\footnote{
 $\x_{\a \dot \a}$ is an anticommuting constant ghost field for translations in spacetime.  The expression $\x \cdot \pa \equiv \x_{\a \dot \a}
\pa^{\a \dot \a}$ is needed to complete the transformations, since the anticommutator of two SUSY transformations is a translation. Note that there is a term linking each source and its field, by a translation, as in  
$ U_{R  \dot \a}
 \x \cdot \pa\;
 \c_L^{\dot  \a}
$.  This yields exactly zero when we combine it with the variation which comes from (\ref{susypb}):
$
\d \x_{\a \dot \a}
=
C_{\a} \oC_{ \dot \a}
$
which comes from the action term
$\cA_{\rm SUSY}
=
 h^{\a \dot \a}
C_{\a} \oC_{ \dot \a}$.  
Usually we can (and do) just ignore these kinds of terms, since they just compensate for total derivatives in the variations.}.

{\bf \theorange}.\;\addtocounter{orange}{1}
\la{seczinnch}
The above Zinn action can usefully be rewritten so that the derivatives with respect to fields are easy to see:

\be
\cA_{\rm Zinn\;\c\;Form \;2}
\la{chiwithfieldsout}
=\int d^4 x
\c_R^{\dot \a} 
\lt (
-\fr{1}{2} \pa_{\a  \dot \a} 
 \X
  C^{\a}
\ebp
-
 \W_{\a  \dot \a} 
  C^{\a}
+ \x \cdot \pa\;
 U_{L\dot \a}
\rt )
\eb
+
\c_L^{\dot \a}
\lt (
\fr{1}{2} \pa_{\a  \dot \a} 
 \ov\X
  C^{\a}
-
\ov \W_{\g \dot \a} C^{\g}
+ \x \cdot \pa\;
 U_{R\dot \a}
\rt )
\la{firstofchi}
\ee

\be
+
V^{ \a\dot \a}
 \lt (
 \ov K  \oC_{\dot \a}
C_{\a}
-\fr{1}{2}
\pa_{\a}^{\dot \g} U_{R  \dot \g} 
 \oC_{\dot \a}
-\fr{1}{2}
\pa_{\a}^{\dot \g}  U_{R  \dot \a}
 \oC_{\dot \g}
\ebp
-\fr{1}{2}\pa_{\dot\a}^{  \g} \oU_{L    \g} 
 C_{  \a}
-\fr{1}{2}
\pa_{\dot\a}^{ \g}  \oU_{L  \a}
 C_{\g}
+ \x \cdot \pa\;
 \ov \W_{ \a\dot \a}
\rt )
\eb
+
 \oB 
 \lt (
\ov U_{R    \b}
 C^{\b}
-
  U_{L \dot  \b}
  \oC^{  \dot \b}
+ \x \cdot \pa\;
\X
\rt )
\ee

\be
+
\w\lt (
 \pa^{\g \dot \d}
\ov\W_{\g \dot \d}
 + \x \cdot \pa\;
\oK
\rt )
\eb
+
\h 
\lt (
C_{\b} \oC_{\dot \b}
\pa^{\b \dot \b}
\ov \D
+ \x \cdot \pa\;
\oJ
\rt )
\lt.
+
L  
\lt (
\oJ 
+ \x \cdot \pa\;
\D
\rt )
\rt \}
\la{lastofchi}
\ee

This form $\cA_{\rm Zinn\;\c\;Form \;2}
$ has another very nice use, explained in the next section.

{\bf \theorange}.\;\addtocounter{orange}{1}
\la{zinnforphi}
The Zinn Action  for the $\f$ sector can be derived from the action
$\cA_{\rm Zinn\;\c\;Form \;2}
$
in (\ref{chiwithfieldsout}) to 
(\ref{lastofchi})
 above simply by changing the names of fields and Zinns.  Here it is: 

\be
 \cA_{\rm Zinn\;\f}
\la{zinnphi}=\int d^4 x
 Z_{R}^{\dot \a}
\lt (
-\fr{1}{2} \pa_{\a \dot \a} E
C^{\a} 
-
W_{\a \dot \a} C^{\a}
\ebp
+
\x \cdot \pa\;\f_{L  \dot \a} 
\rt ) 
\la{firstofzphi}
\ee
\be
+
 Z_{L}^{\dot \a}
\lt (
\fr{1}{2} \pa_{\a \dot \a} \oE
C^{\a} 
-
\ovW_{\a \dot \a} C^{\a}
+
\x \cdot \pa\;\f_{R  \dot \a} 
\rt ) 
\ee

\be 
+
\S^{\a\dot \a}
\lt (
 \ov \eta'   \oC_{\dot \a}
C_{\a}
-\fr{1}{2}
\pa_{\a}^{\dot \g} \f_{R  \dot \g} 
 \oC_{\dot \a}
\ebp
-\fr{1}{2}
\pa_{\a}^{\dot \g}  \f_{R  \dot \a}
 \oC_{\dot \g}
-\fr{1}{2}\pa_{\dot\a}^{  \g} 
\ov \f_{L    \g} 
 C_{  \a}
\ebp
-\fr{1}{2}
\pa_{\dot\a}^{ \g}  
\ov \f_{L  \a}
 C_{\g}
+ \x \cdot \pa\;
 \ovW_{\a\dot \a}\rt )
\ee

\be 
+
 \ov \U
\lt ( 
 \ov\f_{R \b}
C^{ \b}
-
 \f_{L \dot \b}
\oC^{\dot \b}
+\x \cdot \pa E
\rt ) 
\ee

\be
+
J'
\lt (
\pa_{\g  \dot \d} 
 \ovW^{\g  \dot \d}  
  + \x \cdot \pa \ov \eta' \rt )
\ee

\be
+
\oK'  
\lt (
C_{\b} \oC_{\dot \b}
\pa^{\b \dot \b}
L'
 + \x \cdot \pa\;
\w'
\rt )
\eb
+
\ov \D' 
\lt (
\w' + \x \cdot \pa\;
L'
\rt )
\la{lastofzphi}
\ee
The substitution, to go from 
(\ref{chiwithfieldsout}) to (\ref{zinnphi}) is 
\be
\c_{R}^{\dot \a}
\ra Z_{R}^{\dot \a}
;\; 
\X\ra E, 
\W_{\a \dot \a}
\ra 
W_{\a \dot \a}
;\cdots {\rm etc.} 
\ee

The new fields $L',\w'$ in the above do not do much.  The field $\h'$ plays an important role in equation (\ref{7one}) below.
This generation of the $\f$ Zinn action from the $\c$ Zinn action is a kind of involution, since doing it twice will bring us back to the original $\c$ Zinn action.

It is natural to ask why this happens.  The author has no answer to that interesting question. But it works nicely as we will see when we look at the spectrum.  It seems to be a form of `BRST Recycling'. If one tries this with the chiral superfield and its Zinn sources, one gets nothing new, because for that case the Zinn sources are also in a chiral multiplet.  But there are many situations where something new will arise.  They need to be examined. 

{\bf \theorange}.\;\addtocounter{orange}{1}
\la{seczinnph}
The Zinn Action  for the $\f$ sector can also be rewritten so that it is easy to take the derivatives by the fields. This yields:

\be
\cA_{\rm Zinn\;\f\;Form \;2}
\eb
=\int d^4 x
 \f_{R  \dot \a}
 \lt (
-\U \oC^{\dot \a}
-
{\widetilde \S}^{(\dot \a \dot \b)} 
 \oC_{\dot \b}
+ \x \cdot \pa\;
 Z_L^{\dot  \a}
\rt)
\ee

\be
+\int d^4 x \f_{L  \dot \a}
 \lt (
+ \ov \U \oC^{\dot \a}
-
{\widetilde {\ov\S}}^{(\dot \a \dot \b)} 
 \oC_{\dot \b}
+ \x \cdot \pa\;
 Z_R^{\dot  \a}
\rt)
\ee

\be
+
 \int d^4 x E
\lt (
-\fr{1}{2}
 \pa^{\a  \dot \a} 
\ov Z_{L \a} \oC_{\dot\a}
+\fr{1}{2}
 \pa^{\a  \dot \a} 
 Z_{R \dot\a}  C_{\a}
+ \x \cdot \pa\;
\ov \U
\rt )
\la{Evar}
\ee

\be
+
\int d^4 x 
W_{ \a\dot \a}
\lt (
-\pa^{ \a\dot \a}
\ov J' -  Z_R^{\dot \a}
C^{\a}
-
 \ov Z_L^{\a}\oC^{\dot \a} 
+ \x \cdot \pa\;
\ov \S ^{\a\dot \a}
\rt )
\ee

\be
+\int d^4 x 
  \h' 
\lt (
- \ov \S^{\b\dot \b}
C_{\b} \oC_{\dot \b}
+ \x \cdot \pa\;
\ov J'
\rt )
\ee

\be
+\int d^4 x 
\ov \w'  
\lt (
-\D' + \x \cdot \pa\;
K'
\rt )
\eb
+\int d^4 x 
\ov L' 
\lt (-
C_{\b} \oC_{\dot \b}
\pa^{\b \dot \b}
K'
+ \x \cdot \pa\;
\D'
\rt )
\ee

In the above we define
\[
{\widetilde \S}^{(\dot \a \dot \b)} 
=
\fr{1}{2}
\lt (
\pa_{\g}^{\dot \a} \S^{\g \dot \b} 
+
\pa_{\g}^{\dot \b} \S^{\g \dot \a} 
\rt )
\]
\be 
{\widetilde {\ov\S}}^{(\a \b)} 
=
\fr{1}{2}
\lt (
\pa_{\dot\g}^{ \a} \ov \S^{\dot\g \b} 
+
\pa_{\dot\g}^{ \b} \ov \S^{\dot\g \a} 
\rt )
\ee

\[
{\widetilde {\ov\S}}^{(\dot \a \dot \b)} 
=
\fr{1}{2}
\lt (
\pa_{\g}^{\dot \a} \ov \S^{\g \dot \b} 
+
\pa_{\g}^{\dot \b} \ov \S^{\g \dot \a} 
\rt )
\]
\be
{\widetilde {\S}}^{(\a \b)} 
=
\fr{1}{2}
\lt (
\pa_{\dot\g}^{ \a} \S^{\dot\g \b} 
+
\pa_{\dot\g}^{ \b} \S^{\dot\g \a} 
\rt )
\ee

Of course the above necessarily is the same as what we started with above in $
 \cA_{\rm Zinn\;\c}$ in equation 
(\ref{chiwithzinnsout}), provided one changes the names of the fields and Zinns appropriately.

{\bf \theorange}.\;\addtocounter{orange}{1}
Now we have written down the kinetic terms and the Zinn terms.  As noted above the two kinetic actions are invariant under separate field transformations.  But now we want a mass term that mixes them.  At this point it is not clear whether one exists.  But it does\footnote{The easiest way to find it is using the spectral sequence, which also shows the existence of other interesting terms.  This will be the subject of a sequel paper}.
For now, we can simply write it down: 
\[
\cA_{\rm Mass\;\c\f}
= \int d^4 x \;\lt \{
m \f_{L \dot \a}
  \c_{R}^{ \dot \a}
+
m \ov\f_{R  \a}
\ov\c_{L}^{  \a}
\rt.
\]
\be
\lt.
+m  E \ov B
+m  W_{\a\dot \a}
\ov V^{\a\dot \a}
+m  \eta' 
\ov\w
\rt \}
+ *
\la{massact}\ee
and it is easy to verify that it satisfies 
\be\lt (
\d_{\rm \c\;Fields}+\d_{\rm \f\;Fields}
\rt )\cA_{\rm Mass\;\c\f}
=0
\ee

{\bf \theorange}.\;\addtocounter{orange}{1}
The $\c$ action in equation
(\ref{kinchi}) has gauge invariance, as is evident from the transfomation of $V_{\a \dot \a}$ contained in line 
(\ref{gaugetrans}).
This calls for a gauge-fixing and ghost action $\cA_{\rm GGF}
$, and we choose:

\be
\cA_{\rm GGF}
=
\int d^4 x \;\d_{\rm \c\;Fields} \lt ( \ov\eta 
\lt[
\fr{1}{2 }\pa_{\a \dot \a} V^{\a \dot \a}
+ \fr{g}{4 } L\rt ]
\rt ) 
+ *
\ee
where $\d_{\rm \c\;Fields}$ is
the BRST transformation of the theory that arises from the Zinn actions above.
Here  the gauge parameter 
$g $  can be chosen to be real $g=\og $.
We can integrate out the auxiliary field $L$ by 
completing the quadratic and shifting, which leaves
\be
\cA_{\rm GF}=-\fr{1}{2 g } \int d^4 x \;   
\lt\{
\lt (   \pa_{\a \dot \a} \ovV^{\a \dot \a}
\rt )
\lt (
 \pa_{\b \dot \b} V^{\b \dot \b}
\rt )
\rt \}
\la{AGF}
\ee

The other part is

\[
\cA_{\rm G}
=
\int d^4 x \; 
\lt \{
\ov \eta 
\Box   \w
+ \eta 
\Box \ov  \w
- g 
 \ov \eta 
 C_{\b} \oC_{\dot \b}
\pa^{\b \dot \b}
\h
\rt \}
\]
\[
+\int d^4 x \; 
\lt \{
-\fr{1}{2 } \ov \eta 
\pa_{\a \dot \a}\lt(
   \c_L^{\dot \a}
C^{\a}
+
 \ov\c_R^{\a}\oC^{\dot \a}
\rt) 
\rt.
\]
\be
\lt.
-\fr{1}{2 }  \eta 
\pa_{\a \dot \a}\lt(
   \c_R^{\dot \a}
C^{\a}
+
 \ov\c_L^{\a}\oC^{\dot \a}
\rt )
\rt \}
\la{ghostaction}
\ee

 {\bf \theorange}.\;\addtocounter{orange}{1}
So we have found a field action $\cA_{\rm Fields}$, which is the sum of 
(\ref{kinchi}),
(\ref{actphi}),
(\ref{massact}),
(\ref{AGF})
and (\ref{ghostaction}).
 We want to see what this free massive action says about the equations of motion of the various fields. 

 {\bf \theorange}.\;\addtocounter{orange}{1}
First we   look at the functional derivatives with respect to the scalar
 fields\footnote{For the field equations we always set the Zinn sources to zero of course}.
\be
\fr{\d \cA_{\rm Fields}}{\d 
B}
=
-2 
\oB
+m \ov E =0
\ee

\be
\fr{\d \cA_{\rm Fields}}{\d 
E}
=
-\fr{1}{2}
 \Box \oE 
+m \ov  B=0
\ee

Putting these together yields

\be
\lt (
 \Box - m^2 \rt )
\oE 
=0
\ee

{\bf \theorange}.\;\addtocounter{orange}{1}
For the vector bosons we have:
\be
\fr{\d \cA_{\rm Fields}}{\d 
\ovV_{ \a\dot \a} }
=
  \fr{1}{2 g}   \pa^{\a \dot \a} 
 \pa \cdot V +
\lt ( \Box V^{\a \dot \a}+
\fr{1}{2}\pa^{\a \dot \a} 
\pa \cdot V \rt ) 
+m
  W^{\a\dot \a}
=0\la{firstvec}
\ee

\be
\fr{\d \cA_{\rm Fields}}{\d 
\ovW_{\a\dot \a}}
= 
W^{\a\dot \a}
+ m
V^{\a\dot \a}=0
\la{secondvec}
\ee

where
\be
\pa \cdot V
\equiv \pa_{\g \dot \g}
V^{\g \dot \g}
; \pa \cdot W
\equiv \pa_{\g \dot \g}
W^{\g \dot \g}
\la{sdefsfsd}\ee

If we choose the Feynman gauge $g=-1$ then this simplifies to 

\be
\lt ( \Box  
-m^2 \rt )
  V^{\a\dot \a}
=0
\ee
For other gauges things are more complicated in the longitudinal part of $V^{\a\dot \a}$.  This would be more interesting in an interacting model of course.

{\bf \theorange}.\;\addtocounter{orange}{1}
\la{ghostfermieq}
Next we turn to the ghost and fermion fields.  We can easily evaluate the following functional derivatives, which yield the equations of motion for these fields:
\be
\fr{\d \cA_{\rm Fields}}{\d 
\c_{L}^{\dot \a}}
=
\pa_{\a\dot\a}
\ov\c_{L}^{ \a}
-\fr{1}{2}\pa_{\a \dot \a} \ov\eta
 C^{ \a}
-m 
\f_{R \dot \a}=0
\la{1one}
\ee

\be
\fr{\d \cA_{\rm Fields}}{\d 
\ov\c_{R}^{ \a}}
=
\pa_{\a\dot\a}
\c_{R}^{\dot \a}
-\fr{1}{2}\pa_{\a \dot \a} \ov\eta
 \oC^{\dot \a}
-m \ov\f_{L   \a}=0
\la{2one}
\ee

\be
\fr{\d \cA_{\rm Fields}}{\d 
\ov \f_{R}^{  \a}
}
=  
\pa_{\a\dot\a}
\f_{R}^{\dot \a}
-\fr{1}{2}  
\ov \eta'  
C_{ \a}
-m 
\ov \c_{L  \a}
=0
\la{3one}
\ee

\be
\fr{\d \cA_{\rm Fields}}{\d 
 \f_{L}^{\dot  \a}
}
=  
\pa_{\a\dot\a}
\ov \f_{L}^{\a}
-\fr{1}{2}  
\ov \eta'  
\oC_{\dot  \a}
-m  \c_{R\dot  \a}
=0
\la{4one}
\ee

\be
\fr{\d \cA_{\rm Fields}}{\d 
\w }
= - \Box \ov \h - m \ov \h'
\la{7one}
\ee

Now define
\be
m \f'_{R \dot \a}
=
\fr{1}{2}\pa_{\a \dot \a} \ov\eta
 C^{ \a}
+m 
\f_{R \dot \a}
\la{phirightprime}
\ee

\be
m \f'_{L \dot \a}
=
+\fr{1}{2}\pa_{\a \dot \a} \ov\eta
 C^{ \a}
+m 
\f_{L \dot \a}
\la{phileftprime}
\ee

For nonzero m, we can write the above equations in the form

\be
\fr{\d \cA_{\rm Fields}}{\d 
\c_{L}^{\dot \a}}
=
\pa_{\a\dot\a}
\ov\c_{L}^{ \a}
-m 
\f'_{R \dot \a}=0
\la{1two}
\ee

\be
\fr{\d \cA_{\rm Fields}}{\d 
\ov\c_{R}^{ \a}}
=
\pa_{\a\dot\a}
\c_{R}^{\dot \a}
-m \ov\f'_{L   \a}=0
\la{2two}
\ee

\be
\fr{\d \cA_{\rm Fields}}{\d 
\ov \f_{R}^{  \a}
}
=  
\pa_{\a\dot\a}
\f_{R}^{'\dot \a}
-m 
\ov \c_{L  \a}
=0
\la{3two}
\ee

\be
\fr{\d \cA_{\rm Fields}}{\d 
 \f_{L}^{\dot  \a}
}
=  
\pa_{\a\dot\a}
\ov \f_{L}^{'\a}
-m  \c_{R\dot  \a}
=0
\la{4two}
\ee

Then it is easy to derive that
\be
\lt (
-\Box 
+m^2 \rt )
\ov \c_{L}^{  \a}
=0
;\;\lt (
-\Box 
+m^2 \rt )
\c_{R}^{\dot  \a}
=0
\ee
and
\be
\lt (
-\Box 
+m^2 \rt )
\f_{R}^{'\dot  \a}
=0
;\; \lt (
-\Box 
+m^2 \rt )
\ov \f_{L}^{'   \a}
=0
\ee
These fermions $\ov\f_{L}^{'   \a}$ and $
\f_{R}^{'\dot  \a}$ are made partly from the antighost, but the mass is not gauge dependent.

{\bf \theorange}.\;\addtocounter{orange}{1}
\la{ghosteq}
Finally we want to get the mass of the ghost $\w$.  We have 
\be
\fr{\d \cA_{\rm Fields}}{\d   \eta 
  }
\eb
=
\Box   \ov \w
- g 
 C_{\b} \oC_{\dot \b}
\pa^{\b \dot \b}
\ov\h
-\fr{1}{2 } 
\pa_{\a \dot \a}\lt(
   \c_R^{\dot \a}
C^{\a}
+
 \ov\c_L^{\a}\oC^{\dot \a}
\rt) 
\la{5one}
\ee

\be
\fr{\d \cA_{\rm Fields}}{\d 
 \eta'  }
=
\fr{1}{2}
 \lt (
 \ov\f_R^{  \d}
C_{\d}
+ \f_L^{\dot\d}
\oC_{\dot \d}
\rt )
+
m \ov \w
\la{6one}
\ee

Adding these equations, if we choose the Feynman gauge $g=-1$, 
with the second multiplied by $m$,  and using definitions
(\ref{phileftprime})
and (\ref{phirightprime}) 
together with  
(\ref{1two}), (\ref{2two}), (\ref{3two}) and (\ref{4two}) yields:
\be
\lt ( \Box - m^2 \rt )
  \ov \w
=0
\ee

{\bf \theorange}.\;\addtocounter{orange}{1}
So, in the Feynman gauge, there are two Dirac fermions, one complex scalar boson, one complex vector boson and one complex ghost field with its antifield, all of them with mass
$m$.
These are the irreducible components for this supermultiplet.  In other gauges,  the longitudinal part of the vector boson and the ghosts are more complicated, which is the normal state of affairs for a vector boson.  In an interacting model, we would expect to be able to show that the S-matrix is gauge independent.

Note that the gauge symmetry is not broken here, even though the gauge boson is massive.  There are no `Higgs' multiplets needed here, and no gauge symmetry breaking of the U(1) carried by the Superspin $\fr{1}{2}$ multiplet here.

The \tr s for the chiral dotted spinor superfield ${\widehat \f}_{L \dot \a}$ were obtained by the trick of using the Zinn \tr s of the $\c$ sector, and the involution map, to convert them to field transformations of the $\f$ sector.    It was in this way that we discovered that the four original superfields resolve themselves into auxiliaries and fields in such a way as to provide an irreducible action formulation for the \DI. This action has interesting BRST cohomology, as will be discussed in a future paper, using spectral sequences to sort things out.

%\end{document}

\vspace{.2in}

\begin{center}
 {\bf Acknowledgments}
\end{center}
\vspace{.2in}

  I thank  Carlo Becchi, Friedemann Brandt, Cliff Burgess, Philip Candelas, Mike Duff, Rhys  Davies, Pierre Ramond, Kelly Stelle, Peter Scharbach,  Raymond Stora, Xerxes Tata, 
J.C. Taylor and Peter West for stimulating correspondence and conversations.

%\tableofcontents

%\tiny \articlenumber\\ \today

\begin{thebibliography}{99}

%\cite{Gatesfundremains}
\bibitem{Gatesfundremains} 
  S.~J.~Gates, Jr., W.~D.~Linch, III, J.~Phillips and L.~Rana,
  ``The Fundamental supersymmetry challenge remains,''
  Grav.\ Cosmol.\  {\bf 8}, 96 (2002)
  [hep-th/0109109].
  %%CITATION = HEP-TH/0109109;%%
  %28 citations counted in INSPIRE as of 19 Feb 2015

%\cite{Gatesfundremains}
%\cite{Calkinsthinkdiff}
\bibitem{Calkinsthinkdiff} 
  M.~Calkins, D.~E.~A.~Gates, Gates, S. James and W.~M.~Golding,
  ``Think Different: Applying the Old Macintosh Mantra to the Computability of the SUSY Auxiliary Field Problem,''
  arXiv:1502.04164 [hep-th].
  %%CITATION = ARXIV:1502.04164;%%

%\cite{Gatesfundremains}
%\cite{Calkinsthinkdiff}
%\cite{Buchbindersupone}
\bibitem{Buchbindersupone} 
  I.~L.~Buchbinder, S.~J.~Gates, Jr., W.~D.~Linch, III and J.~Phillips,
  ``Dynamical superfield theory of free massive superspin-1 multiplet,''
  Phys.\ Lett.\ B {\bf 549}, 229 (2002)
  [hep-th/0207243].
  %%CITATION = HEP-TH/0207243;%%
  %19 citations counted in INSPIRE as of 22 Aug 2014

%\cite{Gatesfundremains}
%\cite{Calkinsthinkdiff}
%\cite{Buchbindersupone}
%\cite{Buchbindersupthree}
\bibitem{Buchbindersupthree} 
  I.~L.~Buchbinder, S.~J.~Gates, Jr., W.~D.~Linch, III and J.~Phillips,
  ``New 4-D, N=1 superfield theory: Model of free massive superspin 3/2 multiplet,''
  Phys.\ Lett.\ B {\bf 535}, 280 (2002)
  [hep-th/0201096].
  %%CITATION = HEP-TH/0201096;%%
  %29 citations counted in INSPIRE as of 20 Feb 2015


%\cite{Gatesfundremains}
%\cite{Calkinsthinkdiff}
%\cite{Buchbindersupone}
%\cite{Buchbindersupthree}
%\cite{KBbook}
\bibitem{KBbook}
Joseph Buchbinder, Sergio M. Kuzenko:
"Ideas and Methods of Supersymmetry and Supergravity"
Chapman \& Hall/CRC, Second Edition,
Series in High Energy Physics, Cosmology and Gravitation,Edition	2, 
ISBN	1584888644, 9781584888642

%\cite{Gatesfundremains}
%\cite{Calkinsthinkdiff}
%\cite{Buchbindersupone}
%\cite{Buchbindersupthree}
%\cite{KBbook}
%\cite{GatesKoutmassless}
\bibitem{GatesKoutmassless} 
  S.~J.~Gates and K.~Koutrolikos,
  ``On 4D, $\mathcal{N} = 1$ massless gauge superfields of arbitrary superhelicity,''
  JHEP {\bf 1406}, 098 (2014).
  %%CITATION = JHEPA,1406,098;%%
  %3 citations counted in INSPIRE as of 22 Feb 2015

%\cite{Gatesfundremains}
%\cite{Calkinsthinkdiff}
%\cite{Buchbindersupone}
%\cite{Buchbindersupthree}
%\cite{KBbook}
%\cite{GatesKoutmassless}
%\cite{Gateslinmassdyn}
\bibitem{Gateslinmassdyn} 
  S.~J.~Gates, Jr. and K.~Koutrolikos,
  ``A dynamical theory for linearized massive superspin 3/2,''
  JHEP {\bf 1403}, 030 (2014)
  [arXiv:1310.7387 [hep-th]].



  \bibitem{Weinberg} Steven Weinberg: ``The Quantum Theory of Fields" Volume 2, Cambridge University Press, ISBN 052155002. 

%\cite{Zinoviev:2007js}
\bibitem{Zinoviev:2007js} 
  Y.~M.~Zinoviev,
  ``Massive N=1 supermultiplets with arbitrary superspins,''
  Nucl.\ Phys.\ B {\bf 785}, 98 (2007)
  [arXiv:0704.1535 [hep-th]].
  %%CITATION = ARXIV:0704.1535;%%
  %9 citations counted in INSPIRE as of 25 Feb 2015



%\cite{Dixonholes}
\bibitem{Dixonholes} 
  J.~A.~Dixon,
  ``Supersymmetry is full of holes,''
  Class.\ Quant.\ Grav.\  {\bf 7}, 1511 (1990).
  %%CITATION = CQGRD,7,1511;%%
  %23 citations counted in INSPIRE as of 20 Feb 2015

%\cite{Dixonchiralcohom}
\bibitem{Dixonchiralcohom} 
  J.~A.~Dixon,
  ``BRS cohomology of the chiral superfield,''
  Commun.\ Math.\ Phys.\  {\bf 140}, 169 (1991).
  %%CITATION = CMPHA,140,169;%%
  %26 citations counted in INSPIRE as of 20 Feb 2015



%\cite{Dixonjumps}
\bibitem{Dixonjumps} 
  J.~A.~Dixon,
  `SUSY Jumps Out of Superspace in the Supersymmetric Standard Model,'
  arXiv:1012.4773 [hep-th].
  %%CITATION = ARXIV:1012.4773;%%
  %1 citations counted in INSPIRE as of 20 Feb 2015




%\cite{Dixonspecseq}
\bibitem{Dixonspecseq} 
  J.~A.~Dixon,
  ``Calculation of BRS cohomology with spectral sequences,''
  Commun.\ Math.\ Phys.\  {\bf 139}, 495 (1991).
  %%CITATION = CMPHA,139,495;%%
  %71 citations counted in INSPIRE as of 20 Feb 2015
 
%\cite{DixonRahm}
\bibitem{DixonRahm} 
  J.~A.~Dixon, R.~Minasian and J.~Rahmfeld,
  ``Higher spin BRS cohomology of supersymmetric chiral matter in D = 4,''
  Commun.\ Math.\ Phys.\  {\bf 171}, 459 (1995)
  [hep-th/9308013].
  %%CITATION = HEP-TH/9308013;%%
  %11 citations counted in INSPIRE as of 20 Feb 2015


%\cite{Dixonholes}
%\cite{Dixonchiralcohom}
%\cite{DixonMin}
%\cite{Dixonjumps}
%\cite{Dixonspecseq}
%\cite{DixonRahm}
%\cite{dix10D}
\bibitem{dix10D}  J.~A.~Dixon,
`Closure of the Algebra and Remarks on BRS Cohomology in D=10 Super Yang-Mills', UTTG Preprint-14-91 (1991) 







 \bibitem{vecsupref} 
For SUSY textbooks, see for example 
 \ci{west,KBbook,superspace,WB}. A useful reprint collection is \ci{ferrarabook}.  

\bibitem{west}  Peter West, Introduction to Supersymmetry and Supergravity, World Scientific (1990).

\bibitem{superspace}  S. J. Gates, M. T. Grisaru, M. Rocek and W. Siegel, Superspace, Benjamin, 1983.

\bibitem{WB}J. Wess and J. Bagger,  Supersymmetry and Supergravity, Second Edition, Princeton University Press (1992).  


\bibitem{ferrarabook} Many of the original papers on  SUSY and supergravity are collected in  Supersymmetry, Vols. 1 and 2, ed. Sergio Ferrara, North Holland, World Scientific, (1987).







 \bibitem{newwest}
A recent textbook is: Peter West,
Introduction to Strings and Branes
,
Cambridge University Press, (2012)
        ISBN: 9780521817479



%\cite{Zinn-Justin:2011bia}
\bibitem{Zinnarticle} 
 A summary and some history can be found in  J.~Zinn-Justin,
  ``From Slavnov-Taylor identities to the ZJ equation,''
  Proc.\ Steklov Inst.\ Math.\  {\bf 272}, 288 (2011).
  %%CITATION = PSMMB,272,288;%%

  \end{thebibliography}
\end{document}